**Honeycomb lattice of graphite probed by scanning tunneling microscopy with a carbon nanotube tip**

 Jeehoon Kim, Junwei Huang, and Alex de Lozanne*

Department of Physics, The University of Texas, TX 78712

*email: delozanne@physics.utexas.edu

A carbon nanotube (CNT) tip was fabricated at the apex of an etched tungsten wire by chemical vapor deposition and used for scanning tunneling microscopy. The honeycomb lattice of graphite in the STM images was resolved with a CNT tip at $T$=79 K. The superior spatial resolution originating from the $p$ orbitals of a CNT is responsible for the image of the honeycomb lattice of graphite in the STM images. The CNT tips are useful to image samples whose lattice constants are small and to get orbital information in samples with orbital ordering due to their superior spatial resolution with the sharp $p$ orbitals.

Low dimensional carbon materials such as one-dimensional (1D) carbon nanotubes (CNTs) and 2D graphene have drawn much attention for decades due to their fascinating physics as well as their high application potential. Intense research on 3D graphite has also been revived following the discovery of quantized magnetoresistance in a single graphene sheet (2D structure)[1] and superconductivity in  graphite intercalated with metal ions[2]. One potential application for CNTs is for use as a tip for scanning probe microscopy (SPM). A CNT tip, grown[3,4] or attached[5] to a conventional cantilever for atomic force microscope (AFM), was demonstrated for use as a high resolution probe[6]: For example, due to its high aspect ratio, a surface with deep and narrow structures was clearly resolved. A metal-coated CNT tip has been also synthesized and used to image nanoscale magnetic structures with high spatial resolution[7,8]. It should be noted that for a scanning probe, the length of a single CNT on the apex of the metallic tip should be less than one micrometer, otherwise vibrational instability will occur while scanning. The advantages of the CNT tip for STM lie in its simple orbital structure (i.e., sharp $p$ orbitals) and its high elasticity, which results in high spatial resolution and immunity from a slight tip crash to the sample surface. Besides their superior electronic and mechanical properties, CNT-STM tips play an important role in the field of a multiple probe STM, which is employed for transport measurements with multiple probes at the nanoscale.[9] The high aspect ratio of the CNT tip makes it possible to bring multiple probes close together (less than 100 nm), which is not possible when using conventional metallic tips. In addition, the CNT-STM tip would resolve the individual orbital shapes that are present in colossal magnetoresistance (CMR) materials due to the sharp $p$ orbital character.

Highly oriented pyrolytic graphite (HOPG) is widely used as a test surface for STM because the atomic image of HOPG is routinely achieved after cleavage at room temperature in air. In HOPG, giant corrugation[10], site asymmetry[11], and super-periodic Moirè patterns[12] have been reported in the STM topography. HOPG has a honeycomb lattice. However, in general, STM only resolves every other atom due to electronic effects, resulting in a triangular lattice with periodicity that is larger than the unit cell distance. The different electronic structure originating from the dissimilarity of the two carbon sites gives rise to the triangular structure in STM, which will be discussed in detail below. In contrast, atomic force microscopy (AFM) shows the honeycomb lattice of HOPG since it directly probes the  van der Waals forces between the tip and the sample, which shows no direct correlation with the density of states of the sample. For instance, Hembacher et al.[13] have observed the honeycomb lattice with their home-built low-temperature STM/AFM employing a turning fork. However, a triangular lattice was also observed in their STM images taken simultaneously with the AFM images. In addition, the honeycomb lattice in HOPG was reported in a single layer of grapene by STM[14] where the carbon atoms are structurally and electronically indistinguishable.

In this paper we report the observation of a honeycomb lattice by STM with a CNT-STM tip. The superior spatial resolution of the CNT tip, originating from the sharp carbon p orbitals, allows the resolution of tunneling currents from the two structurally distinguishable sites, resulting in the honeycomb lattice of HOPG. The CNT-STM tip plays an important role when high resolution imaging is required, for example, for directly imaging of the shape of $d$ orbitals in CMR materials.

All STM measurements were made in a home-built ultra-high vacuum low-temperature STM at $T$=79 K. The HOPG single crystal, available commercially, was cleaved with an adhesive tape in air and immediately transferred into the STM head in the UHV chamber. Chemical vapor deposition (CVD) was used to synthesize CNTs on a tungsten (W) needle that was electrochemically etched. A large radius of curvature of about 500 nm in the W tip is suitable for growing CNTs at the apex. Etched W tips were electroplated with Fe and kept at $T$= 900 K for 10 minutes with flowing helium and ethylene gas inside a quartz tube. After CVD, CNT tips that are suitable for STM were examined inside a scanning electron microscope (SEM). It is worth noting that hydrocarbon molecules originating from a diffusion pump employed in the SEM may contaminate the CNTs during examination. The probability of producing good CNT-STM tips is about 20% in our setup. Since we attempted to synthesize CNT-STM tips with more than five W wires, we routinely produce at least one good CNT-STM tip. To increase the productivity of the CNT tips in a reliable manner we have developed a *in-situ* method to make CNT tips inside the SEM, which is described elsewhere.[5]

Figure 1 shows images of CNTs, grown on the etched tungsten (W) wire, that were taken in scanning electron microscope (SEM). The single CNT, whose length is about 300 nm, was grown at the end of the etched W wire along the tip axis, which is suitable for an STM tip: In general, a CNT with a length more than 1 μm is unfavorable for an STM scanning due to its vibration instability. All STM images in this study were obtained with the CNT tip displayed in Fig. 1. In Fig. 2 we present STM topography of HOPG taken at 79 K with the CNT tip. The field of view in the image is 30 nm x 30 nm where the 512 pixel × 512 pixel exhibits large bright features, marked by dotted circles, as well as individual carbon atoms. Since HOPG has intrinsically many defects between the layers due to its layered structure, the intercalating defects are likely to be responsible for the bright features in the image.

HOPG has a honey comb lattice with two distinct carbon sites (α and β). The stacking sequence between the layers is mismatched laterally by one unit cell distance, resulting in two distinguishable sites: The carbon (C) atom of the α site has a next nearest neighbored C atom along the *c* axis and that of the β site does not. This structural difference also gives rise to the dissimilar electronic structure between them. The α site shows a three dimensional character and its band structure is broad, whereas the electronic band of the β site is localized around the Fermi energy ($E_F$). Theses two different bands selectively play a role, depending on the energy in an STM experiment: i.e., the contribution of the β site to the total tunneling current is more dominant than that of the α site at low bias. This result allows the site selective imaging depending on the polarity of the STM bias. For example, STM only resolves one site out of the two, either α or β atoms depending on the set point bias polarity. As a result, STM sees the triangular lattice instead of the honeycomb lattice. Figure 3 shows a summary of the site selective imaging in STM with the CNT tip. Figure 3(b) shows a triangular lattice of the β atoms at the positive sample bias and the image [see Fig. 3(c)] taken immediately after Fig. 3(b) with the same scanning parameters used in Fig. 3(b) remained unaltered, indicating that the thermal drift was negligible. The image taken at the negative bias [see Fig. 3(d)] shows anti-correlations with those obtained at the positive [see Figs. 3(b)-3(c)], indicative of imaging the α atoms. The energy selective images as demonstrated in Figs. 3(a)-3(d) are in good agreement with those reported previously.[11] Shown in Figure 4 are the time-dependent STM images with the same CNT tip. The time lag between Figs. 4(a) and 4(b) was five days. Interestingly the triangular lattice transformed into the honeycomb lattice, and also the magnitude of corrugation in the images changed from high to low upon scanning for several days. Figures 4(c) and 4(d) show such a transition with the same trend in a large field of view after they are processed for visual clarity. The insets in each image represent raw data.

Several explanations of the honeycomb lattice of HOPG in STM images were previously reported: a double-tip effect[15], a translation of the topmost layer[16], a tip-sample mechanical interaction[17], and an ensemble atom[18]. In particular, Wang *et al.*[19] observed both the triangular and honeycomb structures on HOPG in their STM images. They claimed that the translational sliding of the topmost layer during the sample cleavage was responsible for the observation of the honeycomb structure in the STM image. It is worth noting that all previous explanations invoke a mechanical effect of the tip and/or the sample. Both the triangular and honeycomb lattices, observed in our measurements under the same conditions with the same CNT tip, may have a different origin compared to those found previously. First, we can rule out a mechanical change of the sample and the tip because the scanned area was identical upon transition from triangular to honeycomb and the CNT tip was confirmed in SEM after measurements. Second, the possibility of a dual tip effect can be also excluded because it generally results from the tip-making and -treatment processes by etching and by applying pulses, respectively, which is not the case for the CNT tip. We believe the transition of the lattice structure from triangular to honeycomb found in this study results from the superior resolution of the CNT tip owing to the sharp carbon $p$ orbitals as well as the nanometer size of the CNT tip compared to conventional metallic STM tips.

In particular, the sharp carbon $p$ orbital of the CNT tip is essential to investigate a material system whose lattice constant is small like that of HOPG ($a$ = 1.42 Å) because the wavefunction of $5d$ orbitals is extended and broad compared to that of $2p$ orbitals, and thus the sample with a small lattice constant would be more sensitive to the tip orbital shape. Therefore an STM image depends more on the tip-orbital states when the sample has a small lattice constant. This is prominent in the HOPG sample due to its small lattice constant. For example, Hembacher *et al.*[20] imaged a tungsten tip with the carbon atom of HOPG. The resulting question is why our STM images initially displayed the structure of a triangular lattice instead of a honeycomb lattice. We believe that it is because of contamination of the CNT tip by the SEM process. Prior to the STM measurement, the CNT tip was imaged with an SEM equipped with a diffusion pump, and therefore contamination is possible from hydrocarbon molecules originating from the diffusion pump. A long scan time helps to clean the CNT tip, allowing imaging of a honeycomb lattice of HOPG with a superior resolution, resulting from its sharp $p$ orbital. The tip was also imaged with the SEM after the STM experiments, and no structural changes were discernible. Of course, the SEM is not able to detect changes of a few monolayers, particularly with light elements such as carbon.

In conclusion, the honeycomb lattice of HOPG was resolved by STM with a CNT tip. The superior spatial resolution due to the $p$ orbitals as well as the nanometer size of the CNT tip is responsible for the observed honeycomb lattice in STM. The CNT tips are useful to image samples with small lattice constants where a high spatial resolution is required, and also they may be applied to unveil the shape and characteristic of the $d$ orbitals originating from the orbital ordering in CMR materials.

**Acknowledgements**

The authors thank R. Baumbach for the useful discussions. This work is supported by the National Science Foundation (NIRT: DMR-0404252) and by the Welch Foundation.

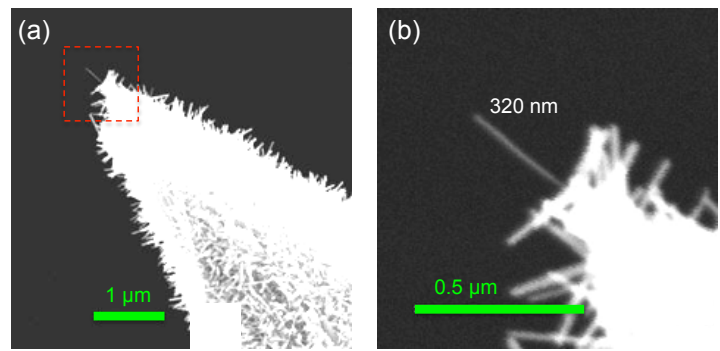

FIG. 1: (Color online) SEM images of carbon nanotubes grown on a tungsten wire by chemical vapor deposition. (a) Single carbon nanotube protruding along the tip axis at the end of the tip that was used for this study. Note that a large number of short CNTs are seen on the side of the etched W tip. (b) The zoom-in image of single CNT. The length of the carbon nanotube on the apex of the tungsten wire is about 300 nm, which is ideal for an STM tip.

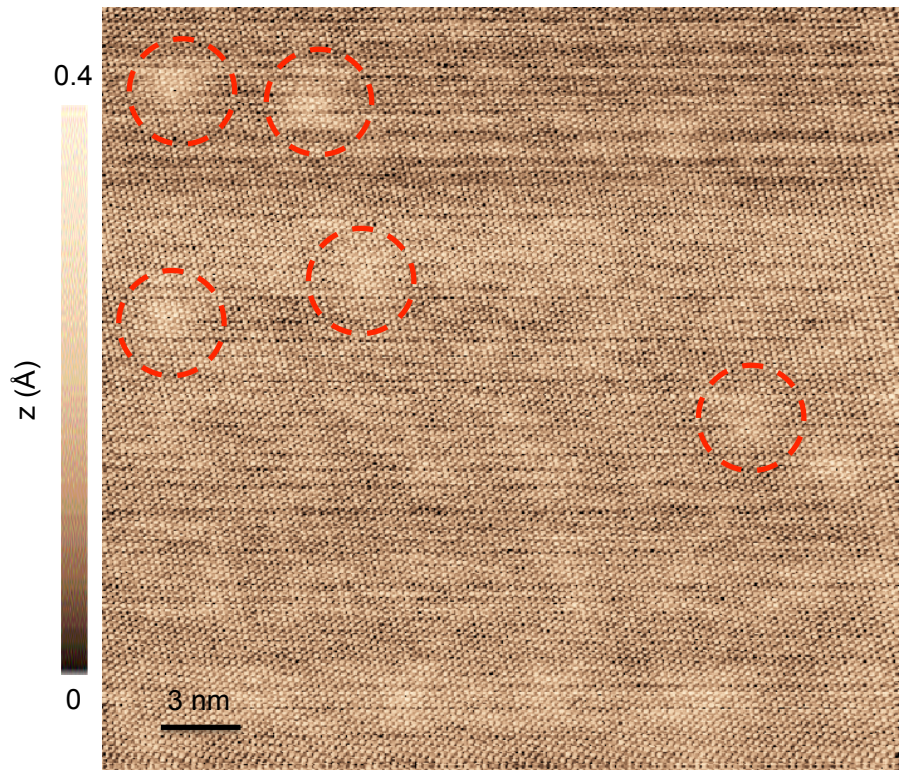

FIG. 2: (Color online) Large field of view obtained from HOPG at $T$=79 K in UHV with the CNT tip shown in Fig. 1. Large bright islands, marked with a dotted red circle, were observed, which may result from intercalating defects due to the layered structure of HOPG. The tunneling junction parameters were a setpoint bias $V_S$ = +0.55 V and a setpoint current $I_S$ = 1 nA, respectively. Individual atoms were clearly resolved in the large field of view. The image frame is made up of $512 \times 512$ pixels for a high spatial resolution.

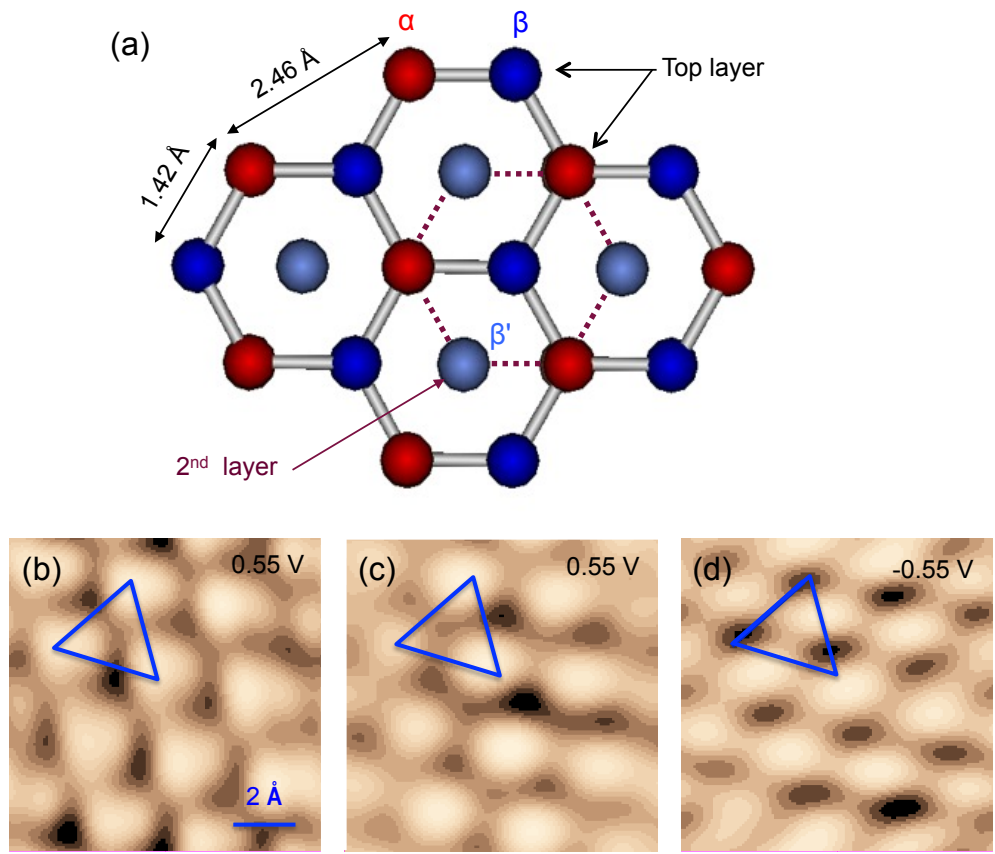

FIG. 3: (Color online) (a) Schematic illustration of a honeycomb lattice in HOPG. Red and blue spheres exhibit α and β atoms, respectively, whose sites are distinguishable owing to a layered structure that is shifted laterally between the top and the second layer by one unit cell distance. (b) Image of β atoms at +0.55 V. (c) Image taken with the same parameters used in (b) and immediately obtained after (b) suggests that a thermal drift is negligible between the images. (d) Image taken at the negative bias of -0.55 V shows anti-correlation compared to the images obtained at 0.55 V.

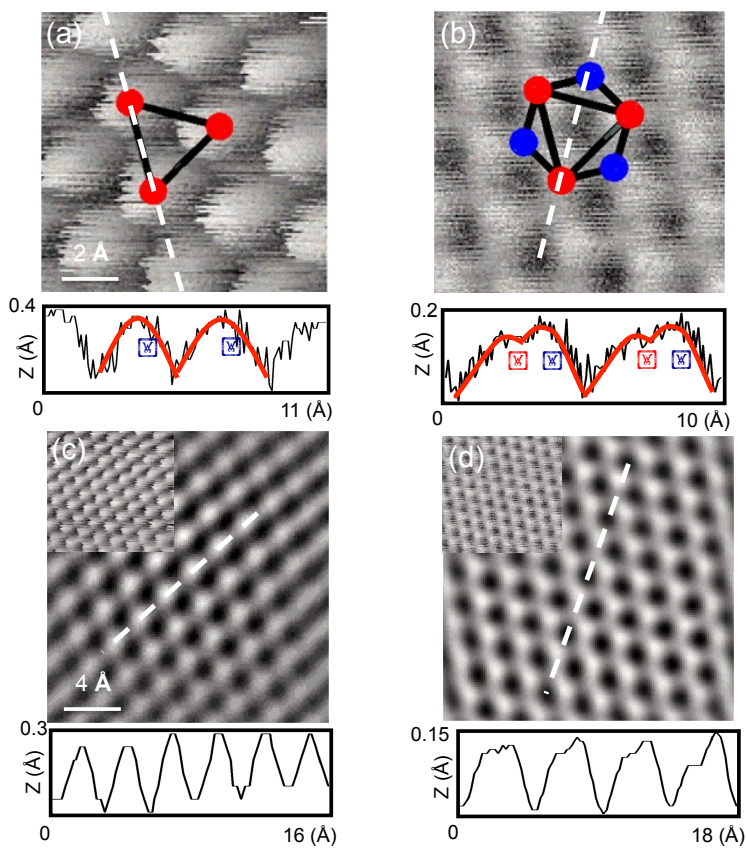

FIG. 4: (Color online) Transition from a triangular lattice to a honeycomb lattice in HOPG (a CNT tip was used). (a) A typical triangular lattice taken with the CNT tip. The vertical corrugation along the dotted line is 0.4 Å. The red spheres represent $\alpha$-site carbon atoms that are visible at a positive sample bias. The line profile is along the dotted line in the image. (b) The honeycomb lattice observed under the same conditions, 5 days after the image in (a) had been taken. Note that the vertical corrugation was reduced down to 0.2 Å. (c) and (d) Processed images obtained with a CNT tip of both a triangular lattice and a honeycomb lattice in HOPG at the same location. The insets show the raw data.